\documentclass[a4paper,preprint,aps]{revtex4}

\usepackage{graphicx}
\usepackage{dcolumn}
\usepackage{bm}

\newcommand{\eqa}{\begin{equation}}
\newcommand{\eqz}{\end{equation}}
\newcommand{\eqma}{\begin{eqnarray}}
\newcommand{\eqmz}{\end{eqnarray}}

\begin{document}
\title{ Assessment of Various Density Functionals and Basis Sets for
the Calculation of Molecular Anharmonic Force Fields }
\author{A. Daniel Boese}
\affiliation{Institute of Nanotechnology, Forschungszentrum
Karlsruhe, P.O. Box 3640, D-76021 Karlsruhe, Germany}
\author{Wim Klopper}
\affiliation{Lehrstuhl f\"ur Theoretische Chemie, Institut f\"ur
Physikalische Chemie, Universit\"at Karlsruhe (TH), D-76128
Karlsruhe, Germany}
\author{Jan M. L. Martin}
\email{comartin@wicc.weizmann.ac.il} \affiliation{Department of
Organic Chemistry, Weizmann Institute of Science, IL-76100
Re\d{h}ovot, Israel}
\date{Received Feb. 14, 2005; Final form Mar. 15, 2005; {\em Int. J. Quantum. Chem.}, in press}
\smallskip
\begin{abstract}
In a previous contribution (Mol. Phys. {\bf 103}, xxxx, 2005), we
established the suitability of density functional theory (DFT) for the
calculation of molecular anharmonic force fields. In the present work,
we have assessed a wide variety of basis sets and exchange-correlation functionals
for harmonic and fundamental frequencies, equilibrium and ground-state rotational constants, and thermodynamic
functions beyond the RRHO (rigid rotor-harmonic oscillator) approximation. The fairly good performance of double-zeta plus polarization basis sets for frequencies results from an error compensation between basis set incompleteness and the intrinsic error of exchange-correlation functionals. Triple-zeta plus polarization basis sets are recommended, with an additional
high-exponent $d$ function on second-row atoms. All conventional hybrid GGA functionals perform about equally well: high-exchange hybrid GGA and meta-GGA functionals designed for kinetics yield poor results, with the exception of of the very recently developed BMK functional which takes a middle position along with the HCTH/407 (second generation GGA) and TPSS (meta-GGA) functionals. MP2 performs similarly to these functionals but is inferior to hybrid GGAs such as B3LYP and B97-1.
\end{abstract}
\maketitle

\section{Introduction}

In recent years, Density Functional Theory (DFT) has perhaps been the most commonly
applied method within the field of computational chemistry.
Harmonic frequencies
can routinely be computed,
and assist experimentalists in the assignment of their measured
infrared spectra.

Several contributions have been made in the last years which have
established the computation of anharmonic force fields more firmly.
The calculation of fundamental frequencies by the use of anharmonic
force fields allows direct comparison between computed and observed
spectroscopic transitions in contrast to harmonic frequencies,
including resonances between several modes as well as overtones.

In the field of computational thermochemistry, quartic force fields
provide both zero-point vibrational energies (ZPVEs) and
thermal corrections beyond the rigid rotor-harmonic oscillator
approximation. In applications where high-temperature data are
important, anharmonic effects will
become non-negligible\cite{h2o}
and the computation of such corrections will become more important.

As early as 1997, DFT was employed for the computation of
anharmonic force fields of polyatomic molecules. Thiel and
coworkers studied three small molecules using several generalized
gradient approximation (GGA) functionals\cite{thiel}. In the
following years, calculations of anharmonic force fields using DFT
have been few and far between: Handy and coworkers have computed the
force fields of ammonia\cite{anharmhandy1},
benzene\cite{anharmhandy2} and diazomethane\cite{Baraille} with
hybrid density functionals.

Starting in 2003, an increasing number of papers reporting DFT
anharmonic force fields have been published and several
groups have been working on this subject. Noteworthy are the
activities in the Handy group, with studies of furan, pyrrole and
thiophene\cite{Rudolf} and phosphorus pentafluoride\cite{Tew}.
Barone\cite{BaroneAzabenzenes} and two of us\cite{Azabenzenes} have independently
published two
detailed studies on the azabenzene series; in the latter paper,
we also explored the
possibility of combining DFT anharmonic force fields with coupled
cluster geometries and harmonic frequencies.
As for validation studies, the group of Hess\cite{Neugebauer}
has published one on a small number of mainly
triatomic molecules. Barone and coworkers have published a series
of articles
\cite{Baroneval,Baroneval2,Baroneval3,Baroneval4,Baroneval5} about
the calculation of anharmonic force fields. Their validation set
consists
of three larger polyatomic molecules (CH$_2$NH, H$_2$CO,
and C$_2$H$_4$) plus
two larger organic molecules
(tetrazine and benzene) to compare harmonic frequencies.

Finally, we have published a detailed study on the calculation of
quartic force fields\cite{anhar1},
derived spectroscopic constants (vibrational
anharmonicities, rotation-vibration coupling constants),
and
thermodynamic functions, all of which were compared
to the corresponding data obtained from accurate ab initio (large basis set coupled cluster)
quartic force fields.
The validation set
included 17 molecules, namely C$_2$H$_2$\cite{c2h2},
C$_2$H$_4$\cite{c2h4}, CCl$_2$, CF$_2$\cite{cf2},
CH$_2$NH\cite{ch2nh}, CH$_2$\cite{anhar1}, CH$_4$\cite{ch4},
H$_2$CO\cite{h2co}, H$_2$O\cite{h2o}, H$_2$S\cite{h2s},
HCN\cite{anhar1}, N$_2$O\cite{n2o}, NH$_2$\cite{nh2}, PH$_3$\cite{ph3},
SiF$_4$\cite{sif4}, SiH$_4$\cite{sih4}, and SO$_2$\cite{so2}:
the citations accompanying each molecule point to the source for the
{\it ab initio} reference data.
In that same paper, we also considered (for a subset of molecules)
convergence in terms of the integration grids used for
energy+gradient and for the CPKS (coupled perturbed Kohn-Sham) step,
as well as in terms of the step size employed for the numerical
differentiation.
Finally, we assessed
the performance of various GGA
and hybrid GGA exchange-correlation functionals,
specifically BLYP\cite{BLYP}, HCTH407\cite{HCTH407}, and PBE \cite{PBE}
among the former, and B3LYP\cite{B3LYP}, B97-1\cite{HCTH93},
B97-2\cite{B97-2}, and PBE0\cite{PBE0}
among the latter.

Briefly summarizing our findings,
we found the computed anharmonicities to be
very sensitive to the DFT integration grid and step sizes. We were
only able to assess a few different basis sets, and concluded that
the TZ2P basis set was likely to be sufficient in all cases.
For basis sets of double-zeta quality, the basis set error was on
the order of the error inherent in the exchange-correlation functionals themselves,
hence results using these basis
sets are to be taken with some caution.
HCTH/407 appeared to
be the most suitable GGA functional for the purpose, and B97-1 the
most suitable hybrid GGA functional, immediately followed by B3LYP.
Somewhat surprisingly, when combining DFT anharmonicities with large
basis set CCSD(T) geometries and harmonic frequencies, GGA
functionals yielded better results than their hybrid counterparts.
In many cases,
ZPVEs (zero-point vibrational energies) and thermal corrections of sufficiently
high quality to be used in conjunction with nonempirical extrapolation-based
computational thermochemistry methods like
W1, W2, and W3 theory\cite{W2,W3} --- or explicitly correlated methods
like CC-R12\cite{KlopperReview} --- can only be achieved by combining
large basis set CCSD(T) geometries and harmonic frequencies with
the DFT anharmonicities.

In the present contribution, we shall assess performance of a wider
variety of exchange-correlation functionals, including several
meta-GGA and meta-GGA functionals, as well as some recent
functionals developed with thermochemical kinetics in mind. We will
also include HF and MP2 results for comparison.

In addition, we will consider a much wider variety of basis sets. Suitable step
sizes and DFT grids were established in the previous paper and will not be
reconsidered here.

\section{Computational Details}

All calculations were run on the Intel/Linux farms of the Martin group and of the Faculty of Chemistry at the Weizmann Institute of Science.

Following the approach first proposed by Schneider and
Thiel\cite{rovib4}, a full cubic and a semidiagonal quartic force
field are obtained by central numerical differentiation (in
rectilinear normal coordinates about the equilibrium geometry) of
analytical second derivatives. The latter were obtained by means of
locally modified versions of
{\sc gaussian~03}\cite{g03};
modified routines from {\sc cadpac}\cite{Cadpac} were used as the
driver for the numerical differentiation routine. In this approach,
the potential energy surface is expanded through quartic terms at
the global minimum geometry like:
\begin{eqnarray}
V=\frac{1}{2}\sum_{i}\omega_i q^2_i +
\frac{1}{6}\sum_{ijk}\phi_{ijk}q_i q_j q_k +
\frac{1}{24}\sum_{ijk}\phi_{ijkk}q_i q_j q_k q_l
\end{eqnarray}
where the $q_i$ are dimensionless rectangular normal coordinates,
$\omega_i$ are harmonic frequencies, and $\phi_{ijk}$ and
$\phi_{ijkl}$ third and fourth derivatives with respect to the $q_i$
at the equilibrium geometry.

All the force fields have been analyzed by means of the
{\sc spectro}\cite{Spectro} and by {\sc polyad}\cite{Polyad} rovibrational
perturbation theory programs developed by the Handy group and by Martin,
respectively.

In all cases, when strong Fermi resonances lead to band origins
perturbed more than about 2 cm$^{-1}$ from their second-order
position, the deperturbed values are reported and resonance matrices
diagonalized to obtain the true band origins. Rotational constants
were similarly deperturbed for strong Coriolis resonances.

Thermodynamic functions beyond the harmonic approximation are
obtained by means of the integration of asymptotic series method as
implemented in the NASA PAC99 program of McBride and
Gordon\cite{PAC99}.

The quadrature grids used are Euler-Maclaurin radial
grids\cite{Han93} with 140 points if the molecule includes only
first-row atoms, and 200 points otherwise. As angular grid the
974-point Lebedev grid\cite{Lebedev} was employed.
For consistency with our previous paper\cite{anhar1}, grid pruning was not applied,
although spot-checking suggests that it would not have affected
results in any significant way.
Such grids are commonly denoted
140$\times$974 and 200$\times$974, respectively.
For the CPKS
(coupled-perturbed Kohn-Sham) steps, we used a different,
significantly coarser (75$\times$194) grid. In our previous paper
we showed that this combination of grids results in computed fundamental
frequencies that are numerically precise to 1 cm$^{-1}$ or better.

The numerical step size determined to be optimal in our previous
work\cite{anhar1} was:
\begin{eqnarray}
q_{step}(i)=\hbox{4}\times\sqrt{\frac{\mu}{\hbox{amu}}}\times
\sqrt{\frac{1000~\hbox{cm}^{-1}}{\omega(i)}}
\end{eqnarray}
The steps are done along the unnormalized Cartesian displacement
vector of the mass weighted normal coordinates.

We have furthermore tightened convergence criteria at all stages of
the electronic structure calculations to $10^{-10}$ or better (no
convergence could be achieved with even tighter criteria). No auxiliary basis sets were employed.

\section{Results and discussion}
\subsection{Basis set}

We have considered the following basis sets (see Table~\ref{table0} for detailed primitive and contracted basis set sizes):
\begin{itemize}
\item Dunning's cc-pV$n$Z correlation consistent basis sets (going from
double-zeta to quadruple-zeta quality) for the first
row\cite{PVXZfirst}, and the cc-pV($n$+d)Z basis sets of Wilson,
Peterson, and Dunning\cite{PVX+dZ} for the second row. (The latter include
additional high-exponent $d$ functions, which have been shown to be
important\cite{so2} for spectroscopic constants of molecules in
which a second-row atom is surrounded by one of more highly
electronegative first-row atoms.)
These basis sets were optimized at the CISD (configuration interaction with all single and double excitations) level.
\item The TZ2P\cite{TZ2P} and DZP \cite{DZP} basis sets. (Note that the
TZ2P version used by most groups actually includes a third $d$
function for second-row atoms.)
\item Jensen's pc-$n$ basis sets which have been specifically optimized for
the BLYP functional\cite{Jensen1,Jensen2,Jensen3,Jensen4}.
The pc-1 and pc-2 basis sets are the same size as Dunning's cc-pVDZ and cc-pVTZ basis
sets, respectively, while the pc-3 basis set is of
$5s4p2d1f$ quality on H, $6s5p3d2f1g$ quality on C--F, and $6s5p3d2f1g$ quality on Si--Cl, and
could arguably be described as a quintuple-zeta rather than a quadruple-zeta basis set.
\item Pople's 3-21G\cite{321G1,321G2}, 6-31G\cite{631G1,631G2}, and
6-311G\cite{6311G1,6311G2,McLean} basis sets (the latter two with various
combinations of diffuse and polarisation functions).
The 3-21G and 6-31G basis sets (and their polarized variants) were optimized at the Hartree-Fock level, the 6-311G basis set at the MP2 level.
\item Ahlrichs' TZV basis set\cite{TZV}, which is usually augmented with the polarization functions of Dunning's cc-pVTZ basis set. We
denote this combination TZVPP.
For second-row systems, it is again strongly recommended to include
an additional $d$ function: a basis set satisfying that requirement, and which shall be denoted TZV(P+1)P, was obtained by combining Ahlrichs' TZV with the polarization functions from the cc-pV(T+d)Z basis set\cite{PVX+dZ}.
\end{itemize}
Relative CPU times for the hybrid functionals can be approximately gauged from the basis set sizes in Table~\ref{table0}. In practice, CPU time scaling is a lot less steep than the theoretical $N^4$ thanks to integral screening and various cutoffs, and scaling becomes more favorable as the molecule grows.

Since we are only comparing basis sets, we used two additional molecules
which have not been included into the validation set before, HOO and
CH$_{3}$. Both force fields have previously been computed\cite{CH3,HOO} with great
accuracy by {\it ab initio} methods.
The lowest bending mode ($\Pi_g$) of acetylene requires the addition
of diffuse functions in order to obtain a qualitatively correct
anharmonicity ---  both {\it ab initio}\cite{c2h2} and
DFT\cite{anhar1} --- and hence has been eliminated from
consideration here.
Overall, this leaves us with 96 harmonic/fundamental frequencies in our
validation set.

The most saturated basis set used is, as our
experience shows\cite{BMH2}, the pc-3 basis set (see above).
As the Dunning correlation consistent basis sets were developed for correlated ab initio methods,
exponents for the polarization functions (which in their case are more appropriately named angular correlation functions) are biased to the high-exponent region. As a result, they do not appear to be
the most efficient choice of basis sets for Hartree-Fock or density functional theory, at least
for atomization energies\cite{BMH}.
We compare {\it all} basis sets to the largest
one used. For B3LYP, where we only looked at five basis sets, the
most accurate basis set used is the pc-2 basis set, as the
results from Table~\ref{table1} show.

Like for atomization energies, the 3-21G basis set is yielding
extremely large errors. Its RMS error compared to our largest basis
set is 150 cm$^{-1}$, and even the anharmonic correction has an RMS
error of 30 cm$^{-1}$. Even worse, some of the force fields
yield positive anharmonic corrections. The unpolarized 6-31G basis set
fares little better for harmonic and fundamental frequencies.
However,
unlike the 3-21G basis set, the anharmonic corrections for most molecules
are at least somewhat reasonable. Nevertheless, even here we
experienced (albeit small) positive anharmonic corrections for two
molecules. The first basis set which gives good results is the
6-31G* basis set. The RMS error for the harmonic and fundamental
frequency is still considerable at 46 cm$^{-1}$, however
no more qualitatively wrong anharmonicities are seen.
The rest of the double-zeta basis sets seem to give results which
are comparable to each other(RMS errors around 30 cm$^{-1}$),
whereas the 6-31+G* and pc-1 basis sets are a bit closer to the
basis set limit. Starting with basis sets of triple-zeta quality,
the basis set errors in the harmonic frequencies get significantly
smaller, ranging from 16 to 5 cm$^{-1}$. Pople's basis sets of
triple-zeta quality do not perform well compared to, e.g., the
TZ2P basis set, only the 6-311+G(3df,2pd) with all possible
polarization functions shows
a low error. The 6-31G basis sets also seem to require diffuse
functions to yield results comparable to the other basis sets
double-zeta quality. This reiterates the observation made in our
contribution where we looked at basis set effects for atomization
energies\cite{BMH}. Here, we concluded that ``diffuse functions are
more important with the Pople basis sets, in that they provide a
significant error reduction''. This implies that the 6-31G basis
sets are slightly less balanced than the other used basis sets.
When looking at the basis set error of the TZVPP basis set, the
extra $d$ polarization function lowers the error by another 30\%,
making it the basis set with the lowest error of all
triple-zeta basis sets tested. Finally, this error is again reduced by half
by the only other basis set of quadruple-zeta quality tested,
indicating that the basis set is very close to the Kohn-Sham limit
for these properties.
Exactly the same behavior can be observed for the B3LYP functional,
with similar errors compared to the pc-2 reference basis set. Thus,
we can expect {\it all} hybrid functionals with 20\% exchange to
give similar basis set errors. Interestingly, although the intrinsic
basis set errors of different basis sets look like they are the same
at the triple-zeta level, compared to the pc-2 basis set the error
of e.g. 6-311+G(3df,2pd) basis set is still 8.3 cm$^{-1}$). This
implies that the errors for the different triple-zeta basis
sets do not seem to be consistent. Relative to pc-3, the average basis
set incompleteness error for DFT harmonic or fundamental frequencies
obeys the following sequence:\\

\noindent cc-pVQZ $\ll$ 6-311+G(3df,2pd) $\approx$ TZV(P+1)P
$\approx$ pc-2 $\approx$ TZVPP $\approx$ cc-pVTZ $<$ TZ2P $<$
6-311+G(3d,2p) $<$ 6-311+G** $\ll$ pc-1 $\approx$ 6-31+G** $<$ DZP
$\approx$ cc-pVDZ $\ll$ 6-31G* $\ll$
6-31G $\ll$ 3-21G\\

\begin{figure}
\includegraphics[width=12cm,angle=270]{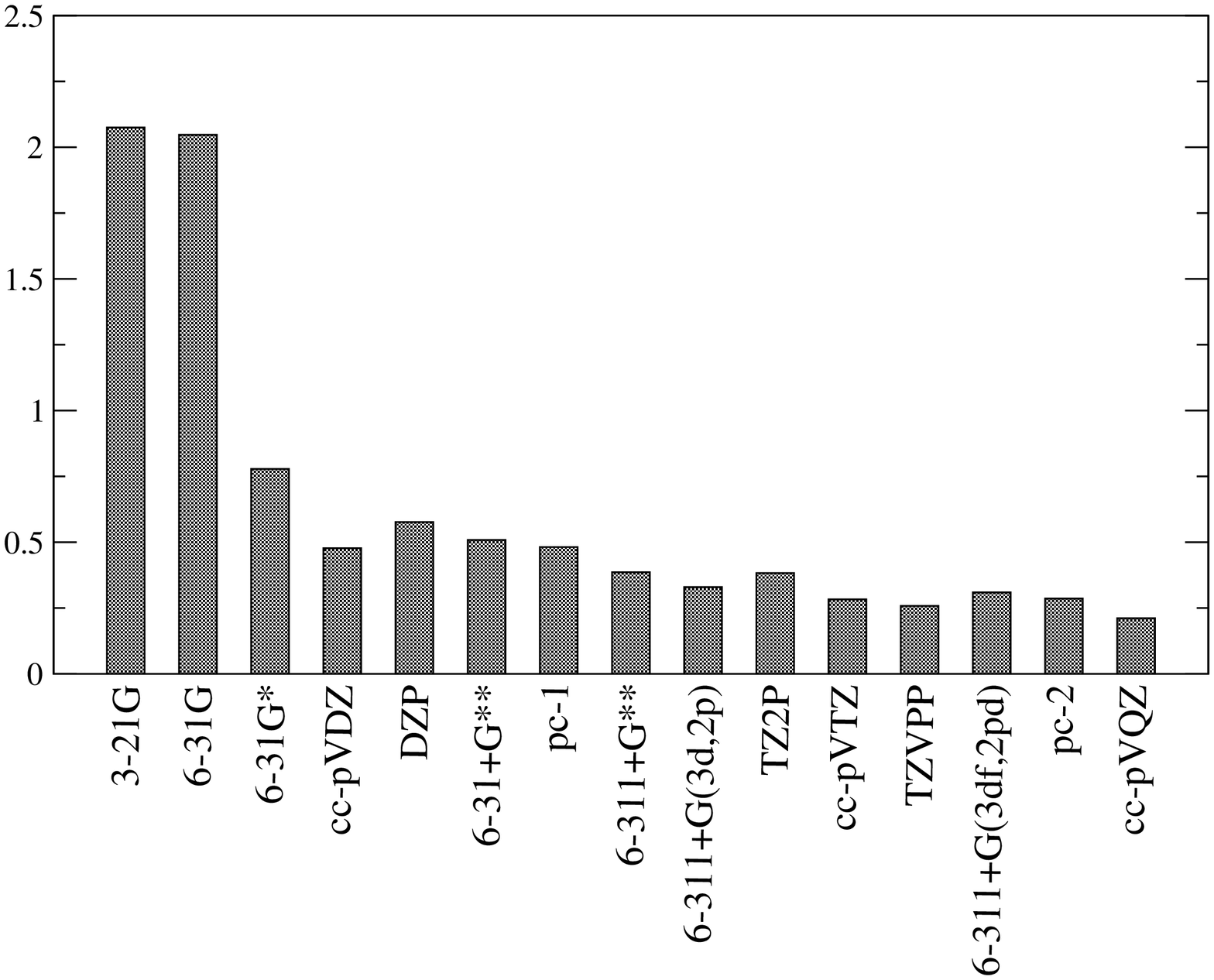} \\
\caption{\label{Figure1}Boese et al}
\end{figure}

\noindent Of course, the error in the fundamental frequencies transfers to
basis set errors in the zero-point energies, as shown in Figure
\ref{Figure1}. Here, considerable error cancellation occurs, and the
larger frequencies obtain also a larger weight. For example, the
TZ2P basis set has about the same basis set error as 6-311+G**,
different from what could be deduced from Table~\ref{table1}. The basis
sets of double-zeta quality with polarization functions (excluding
6-31G*) exhibit errors between
0.58 and 0.48 kJ/mol
compared to
the largest basis set used, and the basis sets of triple-zeta
quality between
0.38 and 0.26 kJ/mol.

In Table~\ref{table2}, we compare the results of B97-1 and B3LYP
with the different basis sets to our reference values. Of course,
error cancellation takes place here, and the basis set error of the
large basis sets will
be masked by the greater
error of the functional itself. Nevertheless, from
this perspective, the results are a bit surprising: For harmonic
frequencies, the TZ2P basis set (which B97-1 has been fitted to) and
the DZP basis set(!) show the lowest error in our validation study.
For B3LYP, the TZ2P basis set
yields the lowest errors as well. Overall, there is little difference
between the different basis sets. Only the 3-21G, 6-31G, cc-pVDZ and
perhaps the 6-31G* basis sets exhibit really large errors. For all the other
basis sets, basis set error is comparable to the intrinsic error for
the exchange-correlation functional.
Still, it might be
possible that some molecules outside our validation set have a
stronger basis set dependence, making at least a basis set of
triple-zeta quality necessary. For example, the aforementioned
C$_2$H$_2$ molecule has a strong basis set dependence and showed
large errors and a positive anharmonic correction when not using
diffuse functions in the basis set. Consistently for all basis sets,
B3LYP has a larger RMS error
than B97-1 in the harmonic frequencies by about 2 cm$^{-1}$.
On the other hand, the anharmonic
correction seems to be described better by the B3LYP functional for
most basis sets (with exception of the pc-1 basis set, which yields
particularly low errors for the anharmonic correction in conjunction
with the B97-1 basis set). Nevertheless, the differences are very
small, and might well change when larger and other molecules would
be considered in the validation set.

\begin{figure}
\includegraphics[width=12cm,angle=270]{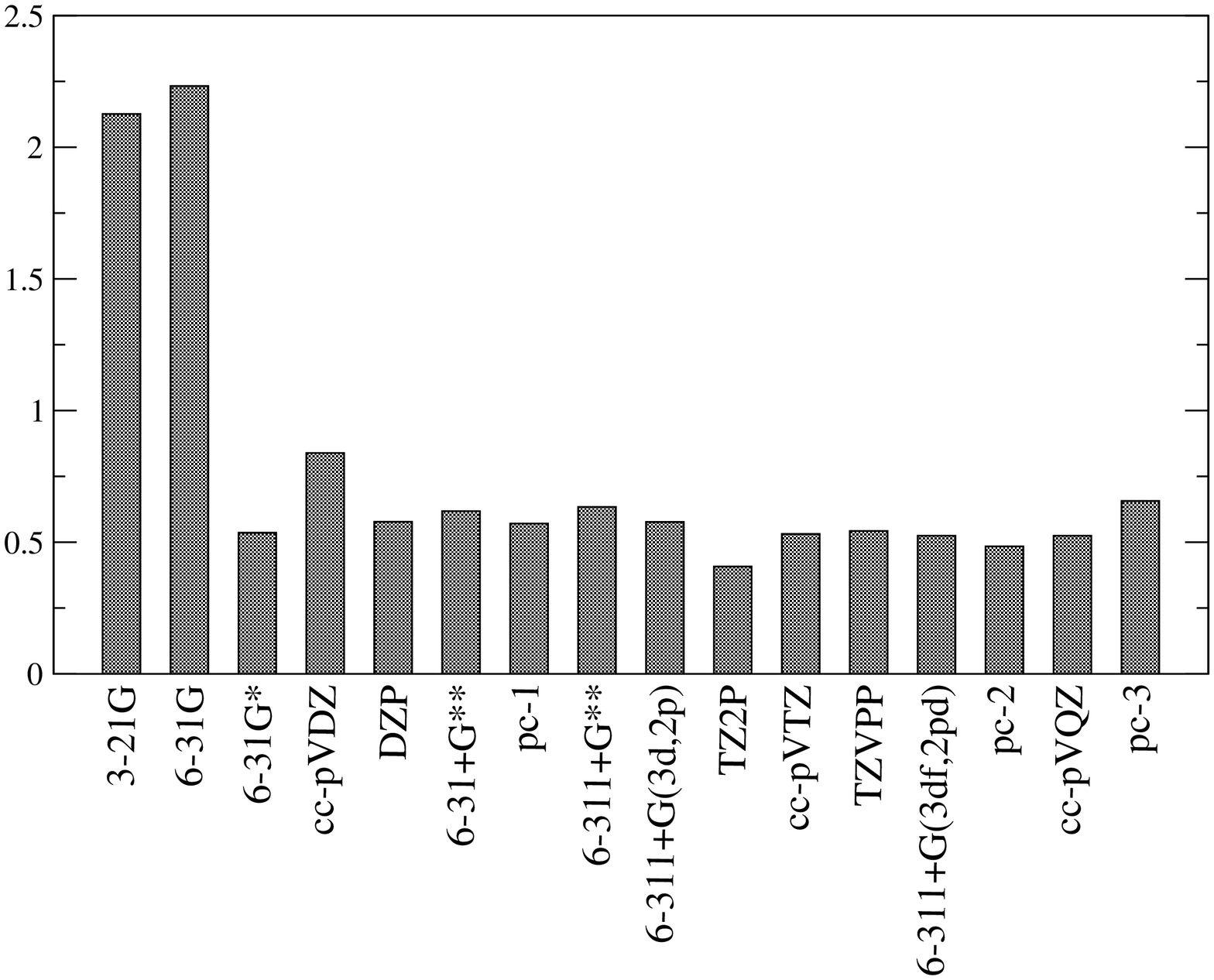} \\
\caption{\label{Figure2}Boese et al}
\end{figure}

We again consider the anharmonic zero-point energies, error statistics for which are
shown in Figure \ref{Figure2}. Like for the fundamental frequencies,
no difference between the basis sets of double and triple-zeta
quality can be seen. The TZ2P basis set once more exhibits the most pronounced
error cancellation between the basis set error and intrinsic error of
the functional.

The basis set errors for the rotational constants (relative to the
largest basis set considered, pc-3) are shown in Table
\ref{table3}. For the TZ2P basis set, the CH$_2$ molecule has been
excluded. The $B_e$ values correlate directly to the accuracy of the
geometry of the molecule. The difference between the errors of the
different $B_e$ and $B_0$ values is quite small, since only the
third derivatives contribute to it. As one might expect, {\em all}
basis sets underestimate the rotational constants (or, conversely,
overestimate bond lengths) relative to the basis set limit.
As for the harmonic frequencies, a very large difference is
encountered when going from the 3-21G and 6-31G basis sets to those
including polarization functions. While the error is cut down to a
third by including diffuse functions for harmonic frequencies,
for rotational constants it is reduced even further, to
approximately one-quarter. Going from double to triple zeta, the
same trend can be observed, with errors in both frequencies and
rotational constants being cut to approximately one-third.
Even without the diffuse functions, the 6-31G* basis set yields a
lower error than the pc-1 basis set. Comparing the basis sets of
triple-zeta quality, the same trend can be observed.
TZ2P seems to be a good compromise between quality and computational
cost, as it yields an RMS error lower than the pc-2 basis set
despite not containing any $f$ functions.
For rotational constants and thus geometries, the extra tight $d$
functions for the TZVPP basis set seems to be extremely important:
The RMS error is reduced from 1\% to 0.29\%
just by including this additional function for the second-row
molecules. While the
TZVPP basis set is one of the worst-performers of the triple-zeta
basis sets tested, the TZV(P+1)P basis set is only surpassed in
accuracy by the 6-311+G(3df,2pd) basis set.
The anharmonic corrections exhibit similar basis set convergence
behavior, with the notable exception of the TZVPP basis set, where
quite large deviations are seen.
Again, the basis set dependencies observed for the B97-1 functional
can be transferred to the RMS errors of the B3LYP functional.

The overall errors compared to our reference values in
Table~\ref{table4} all consistently underestimate the rotational
constants when using either the B3LYP or B97-1 functionals. This
implies that, as we have seen above, when getting closer to the
basis set limit our structures (at least for our validation set)
also become better. As a rule of thumb, this is generally the case
for these two most popular hybrid functionals. Hence, unlike the
case of frequencies --- where double-zeta basis sets sometimes yield
closer agreement with experiment than their quadruple-zeta
counterparts because of error compensation with the intrinsic error
of the functional --- larger basis sets always yield better
agreement with the reference data for {\em geometries}, and at least
a polarized triple-zeta basis set is always recommended.
This implies that for our test set, an (by a small amount) larger
admixture of exchange in the hybrid functional would yield better
results. In the next section, such functionals will be
assessed. Again, the effect when going from $r_e$ to $r_0$
geometries is quite small, at least compared to the error made by
DFT. All basis sets, even including the 3-21G basis set, yield the
same error in the corrections.

For the sake of brevity, we will only report errors in the
thermodynamic functions using the different basis sets. All data (Table~\ref{table5})
are relative
to thermodynamic functions obtained from our reference spectroscopic constants by the same integrated asymptotic series procedure.
Also, we will not discuss here any of the
approximations considered in our previous paper, such as the RRHO
approximation with equilibrium geometries and harmonic frequencies,
the RRHO approximation with zero-point average geometries and
fundamental frequencies or using DFT anharmonic force fields
combined with large basis set CCSD(T) geometries and harmonic
frequencies. Hence, only the data obtained from the pure DFT
anharmonic force field will be presented. Generally, the results of
the thermodynamic functions can be understood form the errors of the
different basis sets in Table~\ref{table2} and Table~\ref{table4},
unless error compensation takes place. Unfortunately, this seems to
be the case, and it is very hard to draw conclusions. The
errors of basis sets of double- and triple-zeta quality show
virtually no difference. Only for the entropy, larger deviations
between the different basis sets can be seen. In our previous study,
it proved to be the only of the three variables under investigation
which showed a lower error for the combination of CCSD(T) harmonic
frequencies and geometries with DFT compared to the pure DFT results
at 2000~K. This implies that the error cancellation effects do not
seem to work as well for the entropy than for the enthalpy or the
heat capacity.

\subsection{Exchange-correlation functionals}

Turning towards the assessment of density functionals, we used for
the functionals the TZ2P basis set for all molecules except
C$_2$H$_2$. For the latter molecule, diffuse functions were
necessary, hence the aug-cc-pVTZ basis set was employed in this case.
In our previous study, we tested the BLYP, HCTH/407, PBE, B97-1, B97-2,
and PBE0 functionals. However, in the last two years some
additional functionals have been published, which may replace the
old functionals. In addition, we have computed {\it ab initio} force
fields with the MP2 and HF methods to compare to the functionals
tested.

In our experience, CPU times with Gaussian 03 do not strongly, or very systematically, depend on the exchange-correlation functional, especially as no auxiliary "density fitting" basis sets are used here. With other codes, GGAs will offer a marked advantage over hybrid functionals. 

The main progress made in recent years in functional development was
the inclusion of the kinetic energy density in a couple of
functionals. This has been done first by Becke\cite{B95}, then by
Van Voorhis and Scuseria\cite{VSXC} as well as by Boese and
Handy\cite{tHCTH}. Lately, Perdew and coworkers\cite{TPSS} have also
published a new meta-Generalized Gradient Approximation (meta-GGA)
functional. The resulting functionals were Bc95 (which however was
later disavowed by Becke himself:\cite{B97} "The functional of Part
IV [i.e., Ref.\cite{B95}] is problematic in very weakly bound
systems... and is therefore not recommended. This will be corrected
in future publications."), VSXC, $\tau$-HCTH and its hybrid and TPSS
and its hybrid version.

Several groups realized\cite{baker95,durant96} that typical hybrid
GGA functionals (with about 20\% exact exchange) have significant
problems in describing barrier heights (albeit less so than pure
GGAs). Durant\cite{durant96} serendipitously found that BHLYP (with
50\% exact exchange\cite{BHLYP}) exhibits these problems to a much
lesser degree than typical hybrid GGAs. Truhlar and
coworkers\cite{mPW1K} took an existing modified Perdew-Wang
functional\cite{PW91,Ada98} and optimized the fraction of
Hartree-Fock exchange for optimal reproduction of a representative
set of reaction barrier heights. The resulting mPW1K functional had
42.8\% of Hartree-Fock type exchange.  However, this functional,
like other hybrid GGA functionals with such a large fraction of
`exact' exchange, performs particularly poorly for atomization
energies and ground-state geometries\cite{BMH}. Truhlar and
coworkers later proposed\cite{mPWB1K} two functionals around the
Bc95 correlation functional, mPW1B95 (with 44\% exact exchange) for
equilibrium thermochemistry and mPWB1K (with 31 \% exact exchange)
for kinetics.

Finally, Boese and Handy conjectured that $\tau$ does not merely
simulate exact exchange (as previously argued\cite{VSXC}), but that
it is capable of simulating {\em variable} exact
exchange\cite{tHCTH}, and thus is potentially capable of
substantially reduced errors in equilibrium properties with
high-exact exchange functionals. Figure 1 of Ref.\cite{BMK} proves
the latter to be the case, and offers at least circumstantial
evidence that the $\tau$ terms can `back-correct' for excessive
exact exchange where it is desirable (that is, near equilibrium
geometries). Boese and Martin\cite{BMK} were thus able to construct
a density functional, termed BMK (Boese-Martin for Kinetics) that
{\it accurately reproduces transition states and reaction barrier
heights without compromising other properties}.

Hence, the new functionals tested are the GGA functionals BLYP,
HCTH/407, PBE (from our previous paper), the TPSS meta-GGA
functional, the hybrid functionals (developed for use of
ground-state properties) B3LYP (20 \% exact exchange), B97-1 (21
\%), B97-2 (21 \%), PBE0 (25 \%), TPSSh (10 \%), and mPW1B95 (31 \%)
and the hybrid functionals (developed for use of transition-state
properties) mPW1K (42.8 \%), mPWB1K (44 \%) and BMK (42 \%).

The use of exact local exchange for hybrid functionals improves the
functionals for many properties,\cite{Goerling}  as this method
selects the correct multiplicative potential of the functional.
However, we did not consider using hybrid functionals with local
exact exchange in this study simply because such an
exchange-correlation potential does not change the energy nor its
nuclear derivatives. Hence, all functionals used will yield quite
close results when using Hartree-Fock or local
exchange.\cite{Karasiev}

When analyzing the data in Table~\ref{table6}, HCTH/407 shows the
lowest RMS errors of all GGA and meta-GGA functionals, while for
example BLYP yields errors almost twice as large. Although TPSS is
clearly an improvement over PBE, it is still less accurate than
HCTH/407 for our validation set. Interestingly, the anharmonic
correction is a lot worse when including the kinetic energy density
here,
with TPSS yielding an error almost twice as large as the HCTH/407 or
BLYP functionals.

Turning towards the hybrid functionals, TPSSh is barely an
improvement over PBE0, and B97-1 is the functional with the lowest
errors for harmonic and fundamental frequencies. B97-2 shows the
lowest RMS error of hybrid functionals for the anharmonic
corrections: this may be related to the fact that
exchange-correlation potential points were used in its
parametrization.
This means that the functional development steps include some
information about the PES, since the fit to the
exchange-correlation-potential means that the functional derivative
is evaluated at every gridpoint of the molecule over full space.
mPW1B95, while having errors for atomization energies comparable to
B3LYP\cite{mPWB1K,mPWB1Keval}, has a much larger error, which can be
almost compared to the error of HCTH/407 functional for the
fundamental frequency. BMK yields, as it was reported for a much
larger set of harmonic frequencies\cite{BMK}, an RMS error for both
harmonic and fundamental frequency comparable to HCTH/407 and
mPW1B95. However, it shows a large error for the anharmonic
correction, which is only surpassed by the Hartree-Fock method. This
of course suggests that the potential energy surface has been
changed, as we could have expected from our previous observations.
(For instance, in the case of the H$_2$S molecule, BMK
underestimates both harmonic stretching frequencies, unlike all
other high-exact exchange functionals. However, the anharmonic
corrections are underestimated for this molecules, resulting in
accidental good agreement with experiment for the fundamentals.) In
general, the anharmonic correction is underestimated on average, and
the harmonic frequency overestimated. This corresponds to a
steepening of the potential energy surface near the minimum, which
is effectively what the kinetic energy density terms `did' to get
the atomization energy correct.
Thus, these results give further insight
into the functional.

We also note that the performance of the different density
functionals for {\em anharmonic} frequencies is consistent with what we found for much larger sets of
{\em harmonic} frequencies.\cite{BMK} Finally, Hartree-Fock unsurprisingly
yields extremely large errors for the frequencies, more than three
times as large as the worst functional tested and ten times as large
as the most accurate functional. The anharmonic correction is not
described well either, although the error here is not as large, only
three times as bad as the best functional. MP2 shows errors close to
HCTH/407 or BMK for harmonic and fundamental frequencies, but the
corrections are still not as accurate as with most density
functionals.
To investigate the arguably stronger basis set dependence of the
MP2, method, we additionally used a larger cc-pVTZ basis set for
MP2. The error is reduced by about ten wavenumbers, which is still
worse than all hybrid functionals with a lower HF percentage than
40\%.
Overall, we would rank the errors of the methods under
investigation for harmonic or fundamental frequencies as follows:\\

\noindent B97-1 $\approx$ B97-2 $\approx$ B3LYP $\approx$ PBE0 $\approx$ TPSSh
$<$ mPW1B95 $<$ HCTH/407 $\approx$ MP2 $\approx$ BMK $\approx$ TPSS
$\ll$ mPWB1K $\approx$ mPW1K $\approx$ PBE $<$ BLYP $\ll$ HF\\

\begin{figure}
\includegraphics[width=12cm,angle=270]{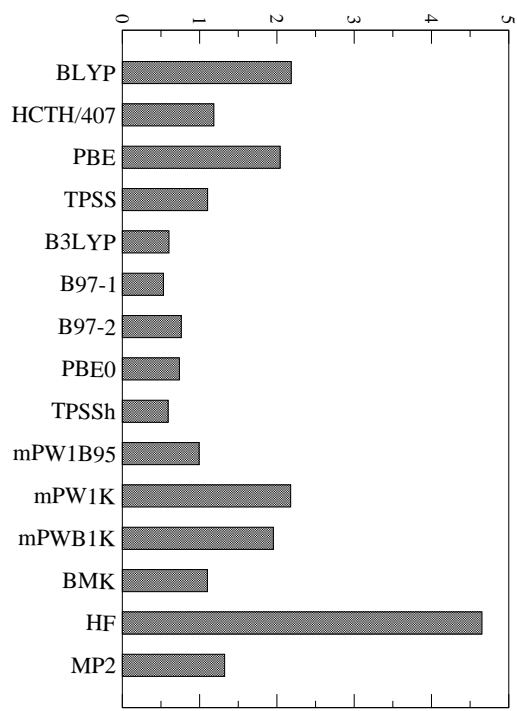} \\
\caption{\label{Figure3}Boese et al}
\end{figure}

\noindent Generally, the errors of the fundamental frequencies transfer to RMS
errors of the zero-point energies of the methods in displayed in
Figure \ref{Figure3}. Again, B97-1 shows the lowest error of all
methods tested, with the error of HCTH/407, TPSS and BMK about twice
as large. Here, MP2 has an RMS error slightly worse than the latter
three functionals. For the computation of accurate transition
states accurate zero-point energies are needed. Usually, a
frequency calculation has to be performed anyhow
(in order to verify that one indeed has found a transition state),
and good performance of the functional for zero-point energies allows one
to `recycle' that frequency calculation for the ZPVE.
Here, we again see
a superior performance of BMK over the other two functionals
developed for such interactions, reducing the error by about half.
This means that the not-so-good performance of mPW1K and mPWB1K is
even worse than expected, since its errors in the zero-point
energies are even close to the worst-performer GGA functionals, BLYP
and PBE.

The errors of the different functionals for rotational constants,
using the TZ2P basis set (again excluding CH$_2$ from the validation
set) are displayed in Table~\ref{table7}. For geometries, the
functionals with $\geq$40\% exchange --- with the notable exception
of BMK --- all do worse than the GGA functionals tested. Both TPSS
and TPSSh yield much poorer results as well. Otherwise, the results
are quite similar to what we observed for the frequencies. HCTH/407
yields the lowest errors of the GGA functionals, and all hybrid
functionals with about 20\% exact exchange yield similar results. By
just ranking our hybrid functionals by the amount of exact exchange,
we can determine their accuracy: TPSSh, with only 10\% exact
exchange yields errors worse than the GGA functional HCTH/407. B3LYP
and B97-1 have 20 and 2 \% exact exchange, respectively. Although
B97-2 also has 21\% exact exchange, it shows lower errors than the
latter two because it has been fit to exchange-correlation potential
points. Finally, PBE0 has 25\% exact exchange, yielding errors for
rotational constants close to B97-2. Nevertheless, TPSS is again an
improvement over PBE, although the improvement is not as large as it
was for harmonic frequencies. For the sake of comparison, MP2 yields
results worse than the `regular exchange' hybrid GGA functionals but
better than most GGA functionals, and also better than the
high-exact exchange functionals (except for BMK).
When increasing the basis set for MP2, the errors are approximately
lowered to those of the hybrid functionals, although it does not
surpass them.
Particularly disappointing is performance for both mPW1K and
mPW1B95. BMK stands out among the high-exact exchange functionals,
yielding RMS errors about half of mPW1K and mPWB1K. Again, the
$B_e-B_0$ corrections are too small to distinguish any functional
from each other. All their errors have about the same values, as it
was the case when comparing the different basis sets in conjunction
with the B97-1 functional to our reference values.
Ranking the functionals for the rotational constants, we arrive at:\\

\noindent PBE0 $\approx$ B97-2 $<$ B97-1 $\approx$ B3LYP $<$ HCTH/407
$\approx$ BMK $\approx$ mPW1B95 $\approx$ MP2 $\ll$ TPSSh $<$ TPSS
$\ll$ mPW1K $\approx$ mPWB1K $<$ PBE $\ll$ BLYP $\ll$ HF\\

The values of three different thermodynamic functions investigated
with different functionals are shown in Table~\ref{table8}. At room
temperature, BMK and MP2 yield the lowest errors of all methods
tested. This is changed at 2000~K; as the higher vibrational levels
become occupied, the problems of both MP2 and especially BMK become
apparent. Since the errors in the anharmonic corrections for both
methods are larger than for 'normal' hybrid functionals, they become
more inaccurate at higher temperatures. Unlike the results from
Table~\ref{table5}, where a lot of error cancellation occurred, most
functionals do perform with errors like the aforementioned rankings,
at least at the medium temperature of 600K. Again, the entropy is
the most sensitive thermodynamic function to the different methods
used.

\section{Conclusions}

We have assessed a wide variety of basis sets and exchange-correlation functionals
for harmonic and fundamental frequencies, rotational constants, and post-RRHO thermodynamic
functions.

For harmonic or fundamental vibrational frequencies, even a double-zeta plus polarization
basis set will only have a basis set incompleteness error comparable to the intrinsic error
in hybrid GGA functionals like B3LYP or B97-1. To be on the safe side, the TZ2P basis set (which in fact is a TZ3P basis set for second-row atoms) can be used. With double-zeta plus polarization basis sets, some error cancellation between basis set and functional errors often occurs.

No such cancellation is seen for geometries or rotational constants: Improving the basis set further is always `worthwhile' here. The TZ2P basis set represents a good compromise between accuracy and computational cost.

No clear conclusions could be drawn for thermodynamic functions, except perhaps for some basis set dependence for entropies at high temperatures.

When using the Ahlrichs TZVPP basis set, performance for second-row molecules can be greatly improved by adding a high-exponent $d$ function, leading to the TZV(P+1)P basis set. In general, a high-exponent $d$
function on second-row atoms is strongly recommended.

For harmonic or fundamental frequencies, there is little to choose between the various `conventional' hybrid GGAs, all of which have intrinsic RMS errors in the 30--35 cm$^{-1}$ range. (We note that our dataset is somewhat biased towards `inorganic' molecules, and better DFT performance can be expected for organic molecules.) Somewhat larger RMS errors in the 50 cm$^{-1}$ range are seen for MP2, HCTH407, TPSS, and BMK. `Kinetics' functionals (except for BMK), as well as first-generation GGAs, yield much poorer results. One downside of BMK are larger errors than GGAs and `conventional' hybrid GGAs for anharmonic corrections. MP2 exhibits the same problem to a lesser extent. Analogous conclusions can be drawn for rotational constants, geometries, and thermodynamic functions.

\section{Acknowledgments}
ADB acknowledges a postdoctoral fellowship from the Feinberg
Graduate School (Weizmann Institute). Research at Weizmann was
supported by the Minerva Foundation, Munich, Germany, by the Lise
Meitner-Minerva Center for Computational Quantum Chemistry (of which
JMLM is a member {\em ad personam}), and by the Helen and Martin Kimmel Center for
Molecular Design. This work is related to Project 2003-024-1-100, "Selected Free Radicals and Critical Intermediates: Thermodynamic Properties from Theory and Experiment," of the International Union of Pure and Applied Chemistry (IUPAC).

 \indent

\newpage

\newpage

\newpage

\clearpage \noindent

Fig. 1:\\
RMS errors (in kJ/mol) for the B97-1 anharmonic ZPVE for several basis sets,
relative to the largest basis
set used.

Fig. 2:\\
RMS errors (in kJ/mol) for the B97-1 anharmonic ZPVE for several basis sets,
relative to the {\it ab initio} reference values.

Fig. 3:\\
RMS errors (in kJ/mol) for several functionals for the anharmonic
ZPVE, compared to the {\it ab initio} reference values.

\newpage
\clearpage

\begin{table}
\caption{Summary of basis sets used in the present work.\label{table0}}
{\footnotesize
\begin{tabular}{l|ccc|ccc|rrr}
\hline\hline
&\multicolumn{3}{c}{Uncontracted}&\multicolumn{3}{c}{Contracted}&\multicolumn{3}{c}{\# basis functions}\\
                & H & 1st row & 2nd row & H & 1st row & 2nd row & H & 1st row & 2nd row \\
\hline
       3-21G    & 3s       & 6s3p        &  9s6p        & 2s     & 3s2p     & 4s3p     &  2 & 9 & 13 \\
       6-31G    & 6s       & 10s4p       & 16s10p       & 2s     & 3s2p     & 4s3p     &  2 & 9 & 13 \\
       6-31G*   & 6s       & 10s4p1d     & 16s10p1d     & 2s     & 3s2p1d   & 4s3p1d   &  2 & 15$^a$& 19$^a$\\
       6-31+G** & 6s1p     & 11s5p1d     & 17s11p1d     & 2s1p   & 4s3p1d   & 4s3p1d   &  5 & 19$^a$& 23$^a$\\
   6-311+G** (b)& 5s1p     & 12s6p1d     & 14s10p1d     & 3s1p   & 5s4p1d   & 7s6p1d   &  6 & 22 & 30 \\
  6-311+G(3d,2p)& 5s2p     & 12s6p3d     & 14s10p3d     & 3s2p   & 5s4p3d   & 7s6p3d   &  9 & 32 & 40 \\
6-311+G(3df,2pd)& 5s2p1d   & 12s6p3d1f   & 14s10p3d1f   & 3s2p1d & 5s4p3d1f & 7s6p3d1f & 14 & 39 & 47 \\
       DZP      & 4s1p     & 9s5p1d      & 11s7p1d      & 2s1p   & 4s2p1d   & 6s4p1d   &  5 & 16 & 24 \\
       TZ2P     & 5s2p     & 10s6p2d     & 12s9p3d      & 3s2p   & 5s4p2d   & 9s6p3d   &  9 & 27 & 42 \\
       cc-pVDZ  & 4s1p     &  9s4p1d     & 12s8p1d      & 2s1p     & 3s2p1d     & 4s3p1d     &  5 & 14 & 18 \\
       cc-pVTZ  & 5s2p1d   & 10s5p2d1f   & 15s9p2d1f    & 3s2p1d   & 4s3p2d1f   & 5s4p2d1f   & 14 & 30 & 34 \\
       cc-pVQZ  & 6s3p2d1f & 12s6p3d2f1g & 16s11p3d2f1g & 4s3p2d1f & 5s4p3d2f1g & 6s5p3d2f1g & 30 & 55 & 59 \\
   cc-pV(D+d)Z  & 4s1p     &  9s4p1d     & 12s8p2d      & 2s1p     & 3s2p1d     & 4s3p2d     &  5 & 14 & 23 \\
   cc-pV(T+d)Z  & 5s2p1d   & 10s5p2d1f   & 15s9p3d1f    & 3s2p1d   & 4s3p2d1f   & 5s4p3d1f   & 14 & 30 & 39 \\
   cc-pV(Q+d)Z  & 6s3p2d1f & 12s6p3d2f1g & 16s11p4d2f1g & 4s3p2d1f & 5s4p3d2f1g & 6s5p4d2f1g & 30 & 55 & 64 \\
       pc-1     & 4s1p     &  7s4p1d     & 11s8p1d      & 2s1p     & 3s2p1d     & 4s3p1d     &  5 & 14 & 18 \\
       pc-2     & 6s2p1d   & 10s6p2d1f   & 13s10p2d1f   & 3s2p1d   & 4s3p2d1f   & 5s4p2d1f   & 14 & 30 & 34 \\
       pc-3     & 9s4p2d1f & 14s9p4d2f1g & 17s13p4d2f1g & 5s4p2d1f & 6s5p4d2f1g & 6s5p4d2f1g & 34 & 64 & 64 \\
       TZVPP    & 5s2p1d   & 11s6p2d1f   & 14s9p2d1f    & 3s2p1d   & 5s3p2d1f   & 5s4p2d1f   & 14 & 31 & 34 \\
       TZV(P+1)P& 5s2p1d   & 11s6p2d1f   & 14s9p3d1f    & 3s2p1d   & 5s3p2d1f   & 5s4p3d1f   & 14 & 31 & 39 \\
\hline\hline
\end{tabular}

\noindent (a) Using Cartesian d functions by default

\noindent (b) 2nd row basis set is actually McLean-Chandler\cite{McLean}
}
\end{table}

\newpage
\clearpage
\begin{table}
\caption{Basis set incompleteness errors (cm$^{-1}$, relative to the pc-3
basis set for B97-1 and to the pc-2 basis set for B3LYP) in harmonic
frequencies, fundamental frequencies, and anharmonic corrections for
several basis sets used in conjunction with the B97-1 and B3LYP
functionals.\label{table1}}
\begin{tabular}{|c c|c c|c c|c c|} \hline\hline
\multicolumn{2}{|c|}{ Property } &\multicolumn{2}{c|}{ Harmonic
Frequency }&\multicolumn{2}{c|}{ Fundamental Frequency }
&\multicolumn{2}{c|}{ Correction}
\\\hline

            & Basis set & mean & RMS & mean & RMS & mean & RMS
\\\hline
\multicolumn{8}{|c|}{ B97-1}\\\hline minimal     & 3-21G    &-31.3
&148.1&-26.2 &141.1& -3.2 & 30.0
\\\hline
double-zeta & 6-31G    &-31.8 &121.9&-30.1 &116.9&  0.6 & 11.6 \\
            & 6-31G*   & 12.9 & 46.6& 14.3 & 46.6&  0.4 &  9.8 \\
            & cc-pVDZ  & -9.3 & 30.7& -7.4 & 31.5& -0.1 &  8.9 \\
            & DZP      &  2.0 & 29.8&  3.5 & 27.3&  0.3 &  5.2 \\
            & 6-31+G** &  3.1 & 26.5&  4.2 & 26.3&  0.7 &  4.5 \\
            & pc-1     & -1.0 & 26.3& -0.1 & 27.3&  0.8 &  6.8 \\\hline
triple-zeta & 6-311+G**& -4.3 & 15.6& -2.3 & 17.1& -0.2 &  3.6 \\
        &6-311+G(3d,2p)& -1.3 & 11.2&  1.0 & 14.4& -0.5 &  3.7 \\
            & TZ2P     &  0.6 &  8.9&  3.0 & 13.0& -0.6 &  3.4 \\
            & cc-pVTZ  & -0.5 &  6.8&  1.9 &  9.7& -0.6 &  2.9 \\
            & TZVPP    & -0.2 &  6.8&  1.2 &  8.9&  0.4 &  2.7 \\
            & pc-2     &  1.3 &  6.3&  2.8 &  8.6&  0.3 &  2.2 \\
    & 6-311+G(3df,2pd) &  0.7 &  5.2&  3.1 &  8.8& -0.5 &  3.1 \\
            & TZV(P+1)P&  1.1 &  4.4&  2.6 &  6.9&  0.3 &  2.8 \\\hline
quadruple-zeta& cc-pVQZ&  0.3 &  2.8&  1.9 &  5.4&  0.2 &  1.5
\\\hline
\multicolumn{8}{|c|}{ B3LYP}\\\hline
 double-zeta & 6-31G*   & 10.1 &45.2&  9.6 & 42.2&  0.5 & 10.0 \\
            & 6-31+G** & -0.1 & 21.1& -0.6 & 20.1&  0.7 &  3.8 \\\hline
triple-zeta & TZ2P     &  0.7 &  9.8&  1.3 &  9.8& -0.6 &  2.5 \\
            & cc-pVTZ  & -1.6 &  9.1& -1.7 & 10.0&  0.1 &  4.6 \\
    & 6-311+G(3df,2pd)  & -0.3 &  8.3& -0.3 &  7.8& -0.1 &  2.5 \\
\hline\hline
\end{tabular}
\end{table}

\newpage
\clearpage
\begin{table}
\caption{Basis set incompleteness errors (cm$^{-1}$, relative to our
reference values) in harmonic frequencies, fundamental frequencies,
and anharmonic corrections for several basis sets used in
conjunction with the B97-1 and B3LYP functionals.\label{table2}}
\begin{tabular}{|c c|c c|c c|c c|} \hline\hline
\multicolumn{2}{|c|}{ Property } &\multicolumn{2}{c|}{ Harmonic
Frequency }&\multicolumn{2}{c|}{ Fundamental Frequency }
&\multicolumn{2}{c|}{ Correction}
\\\hline
Functional  & Basis set & mean & RMS & mean & RMS & mean & RMS
\\\hline
B97-1       & 3-21G    &-38.2 &134.5&-32.4 &127.2& -5.6 & 26.8 \\
            & 6-31G    &-40.7 &108.4&-38.7 &104.6& -2.0 & 11.4 \\
            & 6-31G*   &  4.7 & 36.4&  6.3 & 36.1& -1.4 &  8.3 \\
            & cc-pVDZ  &-17.3 & 44.0&-15.5 & 41.7& -1.7 &  8.0 \\
            & DZP      & -7.8 & 29.6& -5.4 & 29.7& -2.2 &  7.5 \\
            & 6-31+G** & -6.5 & 30.6& -4.7 & 31.3& -1.6 &  8.0 \\
            & pc-1     & -9.2 & 35.0& -7.8 & 34.7& -1.2 &  6.2 \\
            & 6-311+G**&-13.7 & 38.7&-10.8 & 36.3& -2.7 &  9.1 \\
        &6-311+G(3d,2p)&-11.3 & 34.0& -8.3 & 32.6& -2.8 &  8.3 \\
            & TZ2P     & -8.9 & 29.6& -5.7 & 28.3& -3.4 &  8.7 \\
            & cc-pVTZ  & -9.7 & 34.1& -6.5 & 31.2& -3.2 &  8.2 \\
            & TZVPP    & -9.7 & 32.4& -7.5 & 31.1& -2.0 &  8.2 \\
            & pc-2     & -8.2 & 31.5& -5.8 & 30.2& -2.3 &  7.0 \\
    & 6-311+G(3df,2pd) & -8.9 & 33.1& -5.8 & 31.5& -2.9 &  7.7 \\
            & TZV(P+1)P& -8.3 & 31.8& -6.0 & 30.4& -2.1 &  8.2 \\
            & cc-pVQZ  & -9.0 & 32.7& -6.5 & 32.4& -2.4 &  7.8 \\
            & pc-3     & -9.5 & 32.7& -8.7 & 34.0& -2.6 &  7.5 \\\hline
B3LYP       & 6-31G*   &  3.1 & 41.3&  3.6 & 41.3& -0.3 &  6.5 \\
            & 6-31+G** & -8.5 & 34.0& -7.5 & 34.1& -1.0 &  6.8 \\
            & TZ2P     & -7.6 & 31.5& -5.2 & 30.4& -2.3 &  6.7 \\
            & cc-pVTZ  & -9.6 & 36.4& -7.9 & 34.8& -1.7 &  6.2 \\
    & 6-311+G(3df,2pd) & -8.8 & 33.7& -6.6 & 32.8& -2.0 &  6.7 \\
            & pc-2     & -8.3 & 34.5& -6.6 & 33.5& -1.6 &  6.5 \\
\hline\hline
\end{tabular}
\end{table}

\newpage
\clearpage
\begin{table}
\caption{Errors (cm$^{-1}$) for equilibrium and vibrational ground
state rotational constants (and difference between them) for several
basis sets using the B97-1 and B3LYP functional, compared to the
largest basis set used. \label{table3}}
\begin{tabular}{|c c|c c|c c|c c|} \hline\hline
\multicolumn{2}{|c|}{ Property } &\multicolumn{2}{c|}{ $B_e$ (\%) }
&\multicolumn{2}{c|}{ $B_0$ (\%) } &\multicolumn{2}{c|}{ Correction
}
\\\hline
            & Basis set & mean & RMS & mean & RMS & mean & RMS
\\\hline
\multicolumn{8}{|c|}{ B97-1}\\\hline
 minimal    & 3-21G    &-6.13 & 8.66&-6.26 & 8.76& -0.024 & 0.070
\\\hline
double-zeta & 6-31G    &-5.67 & 8.66&-5.72 & 8.70& -0.011 & 0.027 \\
            & DZP      &-1.99 & 2.39&-1.95 & 2.41& -0.003 & 0.014 \\
            & cc-pVDZ  &-1.87 & 2.33&-1.80 & 2.35& -0.008 & 0.032 \\
            & pc-1     &-1.84 & 2.25&-1.80 & 2.27& -0.005 & 0.014 \\
            & 6-31G*   &-1.54 & 1.94&-1.55 & 2.01& -0.012 & 0.038 \\
            & 6-31+G** &-1.43 & 1.78&-1.41 & 1.80& -0.002 & 0.006 \\\hline
triple-zeta & 6-311+G**&-0.93 & 1.25&-0.91 & 1.28&  0.000 & 0.009 \\
            & TZVPP    &-0.47 & 1.00&-0.48 & 1.04& -0.002 & 0.085 \\
            & cc-pVTZ  &-0.39 & 0.75&-0.30 & 1.04& -0.002 & 0.029 \\
            & pc-2     &-0.25 & 0.55&-0.26 & 0.58& -0.001 & 0.002 \\
            & TZ2P     &-0.33 & 0.52&-0.25 & 0.81& -0.005 & 0.027 \\
        &6-311+G(3d,2p)&-0.24 & 0.34&-0.18 & 0.35&  0.000 & 0.005 \\
            & TZV(P+1)P&-0.19 & 0.29&-0.20 & 0.20& -0.002 & 0.085 \\
    & 6-311+G(3df,2pd) &-0.07 & 0.17&-0.08 & 0.18&  0.000 & 0.002 \\\hline
quadruple-zeta& cc-pVQZ&-0.03 & 0.16&-0.05 & 0.17& -0.001 & 0.005
\\\hline
\multicolumn{8}{|c|}{ B3LYP}\\\hline
 double-zeta & 6-31G*  &-1.40 & 1.82&-1.40 & 1.90&-0.012 & 0.042 \\
            & 6-31+G** &-1.30 & 1.52&-1.27 & 1.52&-0.001 & 0.006 \\\hline
triple-zeta & cc-pVTZ  &-0.15 & 0.63&-0.07 & 0.93& 0.002 & 0.029 \\
    & 6-311+G(3df,2pd) & 0.20 & 0.61& 0.21 & 0.61& 0.000 & 0.002 \\
            & TZ2P     &-0.05 & 0.60& 0.12 & 0.48& 0.026 & 0.094 \\
\hline\hline
\end{tabular}
\end{table}

\newpage
\clearpage
\begin{table}
\caption{Errors (cm$^{-1}$) for equilibrium and vibrational ground
state rotational constants (and difference between them) for several
basis sets using the B3LYP and B97-1 functional, compared to our
reference values. \label{table4}}
\begin{tabular}{|c c|c c|c c|c c|} \hline\hline
\multicolumn{2}{|c|}{ Property } &\multicolumn{2}{c|}{ $B_e$ (\%) }
&\multicolumn{2}{c|}{ $B_0$ (\%) } &\multicolumn{2}{c|}{ Correction
}
\\\hline
Functional  & Basis set & mean & RMS & mean & RMS & mean & RMS
\\\hline
B97-1       & 3-21G    &-6.13 & 8.82&-6.22 & 8.87&-0.016 & 0.048 \\
            & 6-31G    &-5.95 & 9.22&-5.96 & 9.25&-0.002 & 0.031 \\
            & DZP      &-2.22 & 2.73&-2.16 & 2.72& 0.004 & 0.038 \\
            & cc-pVDZ  &-2.08 & 2.56&-2.03 & 2.72& 0.000 & 0.032 \\
            & 6-31G*   &-1.75 & 2.21&-1.72 & 2.24&-0.003 & 0.034 \\
            & 6-31+G** &-1.63 & 2.17&-1.57 & 2.23& 0.006 & 0.044 \\
            & pc-1     &-2.02 & 2.59&-1.97 & 2.63& 0.003 & 0.038 \\
            & 6-311+G**&-1.16 & 1.71&-1.57 & 1.80& 0.008 & 0.052 \\
            & TZVPP    &-0.67 & 1.45&-0.60 & 1.58& 0.006 & 0.045 \\
            & cc-pVTZ  &-0.56 & 1.12&-0.37 & 1.09& 0.010 & 0.037 \\
            & pc-2     &-0.41 & 1.07&-0.33 & 1.24& 0.007 & 0.048 \\
            & TZ2P     &-0.46 & 1.01&-0.23 & 1.12& 0.013 & 0.043 \\
        &6-311+G(3d,2p)&-0.42 & 0.88&-0.26 & 1.00& 0.007 & 0.047 \\
            & TZV(P+1)P&-0.35 & 0.84&-0.27 & 1.02& 0.006 & 0.045 \\
    & 6-311+G(3df,2pd) &-0.25 & 0.77&-0.16 & 0.98& 0.007 & 0.048 \\
            & cc-pVQZ  &-0.19 & 0.75&-0.12 & 0.95& 0.006 & 0.045 \\
            & pc-3     &-0.16 & 0.74&-0.07 & 0.97& 0.007 & 0.048 \\\hline
B3LYP       & 6-31G*   &-1.65 & 2.21&-1.64 & 2.27&-0.005 & 0.035 \\
            & 6-31+G** &-1.55 & 2.24&-1.50 & 2.32&-0.005 & 0.045 \\
            & cc-pVTZ  &-0.36 & 0.99&-0.37 & 1.09& 0.008 & 0.037 \\
    & 6-311+G(3df,2pd) &-0.01 & 0.92& 0.06 & 1.17& 0.007 & 0.050 \\
            & TZ2P     &-0.22 & 1.04&-0.23 & 1.12& 0.015 & 0.048 \\
            & pc-2     &-0.20 & 1.17&-0.13 & 1.24& 0.006 & 0.050 \\
\hline\hline
\end{tabular}
\end{table}

\newpage
\clearpage
\begin{table}
\caption{RMS errors using different basis sets for thermodynamic
functions at several temperatures using DFT anharmonic force
fields\label{table5}}
\begin{tabular}{|c c|c c c|c c c|c c c|} \hline\hline
\multicolumn{2}{|c|}{ Property } &\multicolumn{3}{c|}{Heat capacity}
&\multicolumn{3}{c|}{Enthalpy function}
&\multicolumn{3}{c|}{Entropy}
\\\hline
\multicolumn{2}{|c|}{ } &\multicolumn{3}{c|}{ $C_p$ [J/K.mol] }
&\multicolumn{3}{c|}{ $H-H_0$ [kJ/mol] } &\multicolumn{3}{c|}{ $S$
[J/K.mol] }
\\\hline

Functional&Basis set&298.15&600 &2000&298.15&600 &2000&298.15&600 &2000\\\hline
B97-1     & 3-21G   &1.13  &1.03&2.36&0.19  &0.49&2.26& 1.93 &2.57&3.38 \\
          & 6-31G   &1.65  &1.20&1.00&0.35  &0.76&1.50& 3.03 &4.00&4.60 \\
          & 6-31G*  &0.29  &0.36&0.65&0.06  &0.14&0.62& 0.55 &0.67&0.86 \\
          & 6-31+G**&0.50  &0.40&0.76&0.11  &0.24&0.81& 0.84 &1.15&1.53 \\
          &6-311+G**&0.50  &0.40&0.94&0.10  &0.24&0.84& 0.75 &1.06&1.53 \\
          & DZP     &0.56  &0.40&0.72&0.12  &0.26&0.70& 0.96 &1.28&1.54 \\
          & cc-pVDZ &0.57  &0.41&0.57&0.12  &0.27&0.64& 0.98 &1.32&1.56 \\
          & pc-1    &0.46  &0.36&0.57&0.10  &0.22&0.68& 0.80 &1.08&1.41 \\
          & TZVPP   &0.35  &0.29&0.75&0.07  &0.16&0.57& 0.53 &0.74&0.95 \\
          & cc-pVTZ &0.33  &0.30&0.69&0.06  &0.15&0.55& 0.37 &0.56&0.79 \\
          & pc-2    &0.28  &0.23&0.73&0.05  &0.12&0.45& 0.34 &0.50&0.64 \\
          & TZ2P    &0.31  &0.29&0.55&0.04  &0.14&0.56& 0.35 &0.57&0.81 \\
    &6-311+G(3d,2p) &0.42  &0.38&0.80&0.07  &0.19&0.68& 0.45 &0.73&1.09 \\
  &6-311+G(3df,2pd) &0.36  &0.32&0.72&0.06  &0.17&0.58& 0.45 &0.69&0.96 \\
          & cc-pVQZ &0.25  &0.24&0.65&0.05  &0.13&0.44& 0.36 &0.54&0.71 \\
          & pc-3    &0.26  &0.25&0.70&0.05  &0.13&0.57& 0.35 &0.53&0.95 \\
B3LYP     & 6-31G*  &0.34  &0.39&0.51&0.06  &0.16&0.47& 0.54 &0.69&0.81 \\
          & 6-31+G**&0.52  &0.41&0.67&0.11  &0.24&0.70& 0.84 &1.15&1.46 \\
          & cc-pVTZ &0.33  &0.30&0.66&0.06  &0.15&0.49& 0.37 &0.95&1.07 \\
  &6-311+G(3df,2pd) &0.39  &0.35&0.78&0.06  &0.18&0.63& 0.45 &0.70&1.03 \\
          & TZ2P    &0.34  &0.34&0.92&0.05  &0.15&0.76& 0.42 &0.59&0.93 \\
          & pc-2    &0.32  &0.28&0.78&0.05  &0.14&0.49& 0.36 &0.56&0.76 \\
\hline\hline
\end{tabular}
\end{table}

\newpage
\clearpage
\begin{table}
\caption{Errors (cm$^{-1}$) in harmonic frequencies, fundamental
frequencies, and anharmonic corrections for several
methods.\label{table6}}
\begin{tabular}{|c c|c c|c c|c c|} \hline\hline
\multicolumn{2}{|c|}{ Property } &\multicolumn{2}{c|}{ Harmonic
Frequency }&\multicolumn{2}{c|}{ Fundamental Frequency }
&\multicolumn{2}{c|}{ Correction}
\\\hline
Type       & Method  & mean & RMS & mean & RMS & mean & RMS \\\hline
GGA        & BLYP    & -61  & 107 & -60  & 107 & -0.3 &  6.1 \\
GGA        & HCTH/407& -27  &  54 & -26  &  55 & -1.3 &  6.7 \\
GGA        & PBE     & -55  &  91 & -53  &  91 & -2.5 &  9.2 \\
mGGA       & TPSS    & -29  &  61 & -26  &  62 & -2.8 & 11.1 \\
Hybrid(AE) & B3LYP   &  -5  &  34 &  -2  &  34 & -2.5 &  7.5 \\
Hybrid(AE) & B97-1   &  -7  &  32 &  -3  &  32 & -4.0 & 10.3 \\
Hybrid(AE) & B97-2   &   9  &  33 &  11  &  37 & -2.6 &  6.6 \\
Hybrid(AE) & PBE0    &   8  &  34 &  12  &  39 & -3.8 &  9.1 \\
Hybrid(AE) & TPSSh   &  -7  &  35 &  -4  &  35 & -2.7 &  8.6 \\
Hybrid(AE) & mPW1B95 &  20  &  43 &  22  &  51 & -2.1 & 12.8 \\
Hybrid(TS) & mPW1K   &  52  &  89 &  56  &  97 & -4.2 &  9.8 \\
Hybrid(TS) & mPWB1K  &  49  &  86 &  50  &  93 & -0.9 & 14.4 \\
Hybrid(TS) & BMK     &  20  &  58 &  26  &  58 & -5.9 & 21.8 \\
{\it ab initio}& HF  & 105  & 324 & 113  & 329 & -7.4 & 22.5 \\
{\it ab initio}& MP2 &  16  &  56 &  21  &  61 & -5.0 & 12.7 \\
{\it ab initio}& MP2/pVTZ &  16  &  45 &  22  &  53 & -5.1 & 12.8
\\\hline\hline
\end{tabular}
\end{table}

\newpage
\clearpage
\begin{table}
\caption{Errors (cm$^{-1}$) for equilibrium and vibrational ground
state rotational constants (and difference between them) for several
exchange-correlation functionals. \label{table7}}
\begin{tabular}{|c c|c c|c c|c c|} \hline\hline
\multicolumn{2}{|c|}{ Property } &\multicolumn{2}{c|}{ $B_e$ (\%) }
&\multicolumn{2}{c|}{ $B_0$ (\%) } &\multicolumn{2}{c|}{ Correction
}
\\\hline
Type       & Method & mean &  RMS & mean & RMS  & mean & RMS
\\\hline
GGA        &BLYP    &-2.28 & 2.77 &-2.19 & 2.81 & 0.014 & 0.051 \\
GGA        &HCTH/407&-0.61 & 1.25 &-0.50 & 1.30 & 0.005 & 0.040 \\
GGA        &PBE     &-2.01 & 2.32 &-1.84 & 2.28 & 0.018 & 0.057 \\
mGGA       &TPSS    &-1.28 & 1.83 &-1.19 & 1.92 & 0.001 & 0.039 \\
Hybrid(AE) &B3LYP   &-0.18 & 1.05 & 0.14 & 1.12 & 0.015 & 0.047 \\
Hybrid(AE) &B97-1   &-0.42 & 1.01 &-0.19 & 1.03 & 0.012 & 0.043 \\
Hybrid(AE) &B97-2   & 0.34 & 0.92 & 0.61 & 1.15 & 0.015 & 0.047 \\
Hybrid(AE) &PBE0    & 0.25 & 0.90 & 0.59 & 1.19 & 0.018 & 0.051 \\
Hybrid(AE) &TPSSh   &-0.95 & 1.66 &-0.89 & 1.81 & 0.002 & 0.035 \\
Hybrid(AE) &mPW1B95 & 0.90 & 1.30 & 1.10 & 1.64 & 0.009 & 0.064 \\
Hybrid(TS) &mPW1K   & 1.76 & 2.05 & 2.00 & 2.42 & 0.018 & 0.051 \\
Hybrid(TS) &mPWB1K  & 1.93 & 2.19 & 2.30 & 2.66 & 0.018 & 0.051 \\
Hybrid(TS) &BMK     & 0.34 & 1.28 & 0.57 & 1.56 & 0.017 & 0.065 \\
ab initio  &HF      & 3.22 & 3.77 & 3.60 & 4.28 & 0.022 & 0.079 \\
ab initio  &MP2     &-0.05 & 1.35 &-0.02 & 1.46 & 0.005 & 0.042 \\
ab initio  &MP2/pVTZ& 0.12 & 1.06 & 0.23 & 1.07 & 0.013 & 0.102
\\\hline\hline
\end{tabular}
\end{table}

\newpage
\clearpage
\begin{table}
\caption{RMS errors for thermodynamic functions at several
temperatures using DFT anharmonic force fields\label{table8}}
\begin{tabular}{|c c|c c c|c c c|c c c|} \hline\hline
\multicolumn{2}{|c|}{ Property } &\multicolumn{3}{c|}{Heat capacity}
&\multicolumn{3}{c|}{Enthalpy function}
&\multicolumn{3}{c|}{Entropy}
\\\hline
\multicolumn{2}{|c|}{ } &\multicolumn{3}{c|}{ $C_p$ [J/K.mol] }
&\multicolumn{3}{c|}{ $H-H_0$ [kJ/mol] } &\multicolumn{3}{c|}{ $S$
[J/K.mol] }
\\\hline
Type       &Method&298.15&600 &2000&298.15&600 &2000&298.15&600 &2000\\\hline
GGA        &BLYP  &0.94  &0.96&1.09& 0.15 &0.43&1.48& 1.08 &1.69& 2.51 \\
GGA      &HCTH/407&0.64  &0.56&0.62& 0.09 &0.29&0.96& 0.63 &1.07& 1.61 \\
GGA        &PBE   &0.89  &0.92&1.04& 0.14 &0.40&1.27& 0.94 &1.51& 2.19 \\
mGGA       &TPSS  &0.62  &0.46&0.95& 0.11 &0.22&0.89& 0.62 &0.94& 1.37 \\
hybrid(AE) &B3LYP &0.55  &0.39&0.90& 0.08 &0.22&0.79& 0.54 &0.82& 1.14 \\
hybrid(AE) &B97-1 &0.58  &0.47&0.53& 0.09 &0.23&0.62& 0.52 &0.85& 1.11 \\
hybrid(AE) &B97-2 &0.45  &0.41&0.64& 0.06 &0.22&0.71& 0.44 &0.81& 1.17 \\
hybrid(AE) &PBE0  &0.45  &0.41&0.98& 0.07 &0.19&0.94& 0.44 &0.70& 1.23 \\
hybrid(AE) &TPSSh &0.55  &0.45&0.76& 0.10 &0.26&0.75& 0.74 &1.10& 1.46 \\
hybrid(AE)&mPW1B95&0.55  &0.42&0.90& 0.08 &0.15&0.85& 0.47 &0.59& 1.10 \\
hybrid(TS) &mPW1K &0.70  &0.86&0.91& 0.08 &0.29&1.40& 0.60 &1.04& 1.94 \\
hybrid(TS) &mPWB1K&0.55  &0.75&0.91& 0.09 &0.24&1.21& 0.62 &0.94& 1.73 \\
hybrid(TS) &BMK   &0.43  &0.67&1.56& 0.04 &0.22&1.45& 0.26 &0.63& 1.51 \\
ab initio  &HF    &1.60  &2.09&1.62& 0.19 &0.70&3.00& 1.26 &2.32& 4.21 \\
ab initio  &MP2   &0.44  &0.63&0.80& 0.05 &0.20&1.05& 0.34 &0.66& 1.30 \\
\hline\hline
\end{tabular}
\end{table}

\end{document}